\documentclass{ws-ijmpcs}
\usepackage{graphicx}

\usepackage{wasysym}
\usepackage{hyperref}
\hypersetup{colorlinks=true}
\usepackage{multirow}
\usepackage{graphicx}
\usepackage{epstopdf}
\usepackage{graphicx}
\usepackage{dcolumn}
\usepackage{bm}
\usepackage{color}
\usepackage{textpos}
\usepackage{multicol}
\usepackage{subfigure}
\usepackage{lineno}
\newcommand{\pt}{\ensuremath{p_{\rm{T}}}}
\newcommand{\pta}{\ensuremath{p_{\rm T, assoc}}}
\newcommand{\ptt}{\ensuremath{p_{\rm T, trig}}}

\newcommand{\gmom}    {\mbox{${\rm GeV}/c$}}

\newcommand{\tev}     {\mbox{${\rm TeV}$}}

 \newif\ifpdf
 \ifx\pdfoutput\undefined
 \pdffalse
 \else
\pdfoutput=1
 \pdftrue
 \fi
 \ifpdf
 \usepackage{graphicx}
 \usepackage{epstopdf}
 \else
 \usepackage{graphicx}
 \fi

\begin{document}

\markboth{X. Zhu (for the ALICE Collaboration)}
{Two-particle Correlations in pp and Pb-Pb Collisions with ALICE}

%
\catchline{}{}{}{}{}
%

\title{TWO-PARTICLE CORRELATIONS IN pp AND Pb-Pb COLLISIONS WITH ALICE}

\author{XIANGRONG ZHU (for the ALICE Collaboration)}

\address{Key Laboratory of Quark \& Lepton Physics (MOE) and Institute of Particle Physics, Central China Normal University,
Wuhan 430079, China\\
E-mail: xiangrong.zhu@cern.ch
}


\maketitle

\begin{history}
\received{Day Month Year}
\revised{Day Month Year}
\end{history}

\begin{abstract}
The measurement of azimuthal correlations between two particles is a powerful tool to investigate the
properties of strongly-interacting nuclear matter created in ultra-relativistic heavy-ion collisions. We present measurements
of di-hadron correlations in azimuth angle and pseudorapidity in Pb-Pb collisions at $\sqrt{s_{\rm NN}} = 2.76~\tev$ and photon-hadron
correlations in pp collisions at $\sqrt{s} = 7~\tev$ with the ALICE detector, respectively. In di-hadron correlations, 
the near-side jet shape in the short-range correlation region is quantitatively analyzed at $\ptt < 8~\gmom$, 
and the modification of jet-particle yield, $I_{\rm AA}$ and $I_{\rm CP}$, is measured at $8 < \ptt < 15~\gmom$. 
In photon-hadron correlations, isolated leading photon and associated hadrons in its opposite azimuthal direction correlations 
are used to estimate the jet fragmentation function via the imbalance parameter, 
$x_{\rm E} \equiv -\frac{\vec{p}_{\rm T}^{\gamma}\cdot\vec{p}_{\rm T}^{\rm h^{\pm}}}{|\vec{p}_{\rm T}^{\gamma}|^{2}}$.
\end{abstract}


\section{Introduction}
The research objective of ultra-relativistic heavy-ion collisions is to explore the properties of the quark-gluon 
plasma (QGP), a deconfined state of quarks and gluons. Many experimental and theoretical studies of the QGP have been 
obtained from the study of hadron jets, the fragmentation products of high transverse momentum ($\pt$) 
partons\cite{HIP,IArsen1,IArsen1,IArsen2,IArsen3,IArsen4}. It is generally accepted that prior to hadronization, 
partons lose energy in the extreme hot and dense medium due to gluon radiation and multiple collisions. These phenomena
are broadly known as ``jet quenching''\cite{JetQuenching1,JetQuenching2,JetQuenching3}.
At the LHC, the strong jet quenching in central heavy-ion collisions has been reported by ALICE,
ATLAS and CMS collaborations\cite{Raa1,Raa2,Raa3}. The nuclear suppression factor $R_{\rm AA}$, which quantifies the suppression of charged hadrons,
in central Pb-Pb collisions at $\sqrt{s_{\rm NN}} = 2.76~\tev$ is about 0.14 at $\pt\sim7~\gmom$\cite{Raa1,Raa3}. Furthermore,
a strong di-jet energy asymmetry for leading jet transverse momenta above $100~\gmom$ has been reported\cite{jetAsy1,jetAsy2}. At low 
transverse momenta ($p_{\rm T,jet} < 50~\gmom$), background fluctuations due to the underlying event dominate\cite{bkg}
and event-by-event jet reconstruction becomes difficult. Two-particle correlations allow the study of medium effects
on the jet fragmentation without the need for jet reconstruction. 

Especially, direct photon-hadron correlations offer two major advantages as compared to di-jet measurements 
because of the nature of the photon. First, in contrast to partons, photons do not carry color charge and hence
do not interact strongly when traversing the medium\cite{photonwithMedium}. Second, the direct photon 
production at leading order (LO) in pp and A+A collisions is dominated by the QCD compton scattering process, 
$q + g \rightarrow q + \gamma$ and $q + q \rightarrow g + \gamma$ annihilation process, and the photon momentum in the 
center-of-mass frame is exactly balanced by that of the recoil parton. For these reasons, direct photon-hadron 
correlations have been considered as a ``golden channel'' for studying the properties of parton energy loss
including parton fragmentation function without the need of the jet reconstruction\cite{godenChannel1,godenChannel2}.
Furthermore, significant measurements about parton energy loss in the medium by isolated photon-jet correlations at CMS 
are presented in\cite{CMSphoton}.

This proceeding is organized as follows: Sec.~\ref{analysisData} briefly presents the ALICE detector relevant to this
analysis and data sample. Sec.~\ref{twoparticlecorr} shows the method of two-particle correlations. The near-side jet shape
analysis is discussed in Sec.~\ref{nearsidejetshape} and the modification factor of jet-particle yield is discussed in 
Sec.~\ref{modificationfactor}. Sec.~\ref{fragmentationfun} has the discussion of fragmentation function estimation from 
isolated photon-hadron correlations. Sec.~\ref{summary} summarizes the results from this proceeding.

\section{Detector and data sample}
\label{analysisData}
The analyzed data were taken with the ALICE detector described in detail in\cite{ALICEdetector}. The collision vertex finding
and tracking are performed using information from the Inner Tracking System (ITS) and the Time Projection Chamber (TPC). The ITS consists of six 
layers equipped with Slicon Pixel Detector (SPD), Silicon Drift Detector (SDD) and Silicon Strip Detector (SSD). The TPC is a
cylindrical drift detector with uniform acceptance in azimuth angle ($\phi$) and a pseudorapidity coverage of $|\eta| <0.9$. 
The reconstructed vertex information is used to select primary track candidates and constrain the $\pt$ of the track. The
forward scintillators (VZERO) determine the centrality of the Pb-Pb collisions. Details can be found in\cite{ALICEdetails}. 
The photon is detected using the electromagnetic calorimeter (EMCal) which is a Pb-Scintillator sampling calorimeter covering 
$\Delta\phi = 100^{\circ}$ in the azimuthal angle and $|\eta| < 0.7$ in pseudorapidity. 
Photon candidates are selected from energy clusters deposited in the $\pt$ range at 8 to 25 $\gmom$ by photon identification
cuts. The photon identification cuts include track matching, cluster time and the cluster shower shape long axis parameter, 
$\lambda_{0}^{2}$, defined as:

\begin{equation}
 \lambda_{0}^{2}=0.5\times(d_{\eta\eta}+d_{\phi\phi})+\sqrt{0.25\times(d_{\eta\eta}-d_{\phi\phi})^{2}+d_{\eta\phi}^{2}}
\end{equation}
where $d_{ii}$ is the cluster position in $i$ direction weighted by the cell energy.

In the di-hadron correlation analysis, about 14 millions minimum-bias Pb-Pb collision events with an integrated luminosity 
($L_{\rm int}$) of 1.7 $\mu b^{-1}$ at $\sqrt{s_{\rm NN}} = 2.76~\tev$ collected in fall 2010 and 37 millions pp events with
the $L_{\rm int}$ of 6.8 $n b^{-1}$ from March 2011 at $\sqrt{s} = 2.76~\tev$ are used. In the isolated photon-hadron correlation 
analysis, 10 million pp events with the $L_{\rm int}$ of 500 $n b^{-1}$ at $\sqrt{s} = 7~\tev$ triggered by the EMCal 
with a trigger threshold about $5~\gmom$ is used for achieving the measurement of high-$\pt$ photons up to $25~\gmom$ with 
enough rate.

\section{Correlation analysis}
\label{twoparticlecorr}
The $\pt$ dependence of the correlation is studied by measuring triggered correlations. In such an analysis, a particle is 
chosen from a $\pt$ region and called the $trigger~particle$. The so called $associated~particles$ from another $\pt$ region 
are correlated to the trigger particle where $\pta < \ptt$. The associated per-trigger yield is measured as a
function of the azimuthal angle difference $\Delta\phi = \phi_{\rm trig} - \phi_{\rm assoc}$ and pseudorapidity difference 
$\Delta\eta = \eta_{\rm trig} - \eta_{\rm assoc}$:

\begin{equation}
\label{corrpair}
Y(\Delta\phi, \Delta\eta) = \frac{1}{N_{\rm trig}}\frac{dN_{\rm assoc}}{d\Delta\phi d\Delta\eta} 
\end{equation}
where $N_{\rm assoc}$ is the number of particles associated to a number of trigger particles $N_{\rm trig}$. 
This quantity is measured for different ranges of $\ptt$ and $\pta$. 

\begin{figure}
\centering
\includegraphics[scale=0.6]{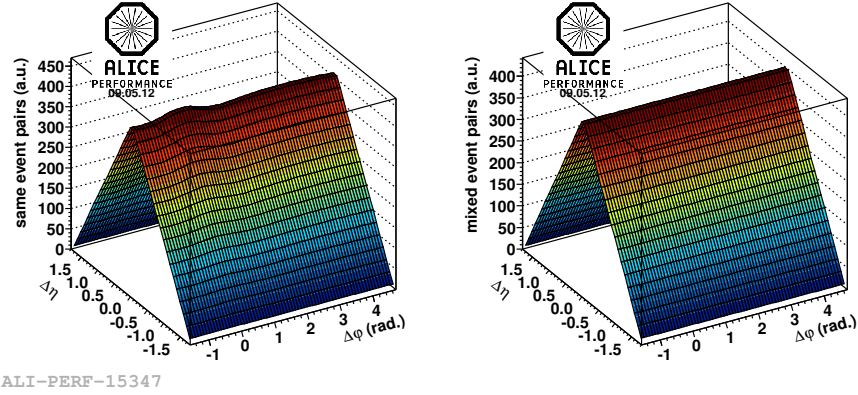}
\caption[]{\label{samemixed}Correlation constructed from pairs of particles from the same events (left panel) and the
mixed events (middle panel)\cite{Jan1}.}
\end{figure}

To obtain the fully corrected per-trigger associated primary particle yield, two steps are performed on the
raw correlations. Firstly, two-track efficiency and acceptance are assessed by using a mixed-event technique: the 
differential yield defined in Eq.~\ref{corrpair} is also constructed for pairs where the trigger and the associated
particle come from different events with similar centrality (or multiplicity in pp) and $z$-vertex position. The angular
correlation constructed from particles within the same event and mixed events are shown in the left and right panel in 
Fig.~\ref{samemixed}. The acceptance corrected distribution can be obtained from the ratio of pair distributions from the same 
and mixed events with a proper normalization factor. The normalization factor is chosen in a way which the distribution
in mixed events is 1 at $\Delta\phi = \Delta\eta = 0$. Secondly, tracking efficiency and track contamination from secondary particles are used to correct 
the correlation function.   
  
  
\subsection{Near-side jet shape}
\label{nearsidejetshape}
\begin{figure}
\includegraphics[scale=0.28]{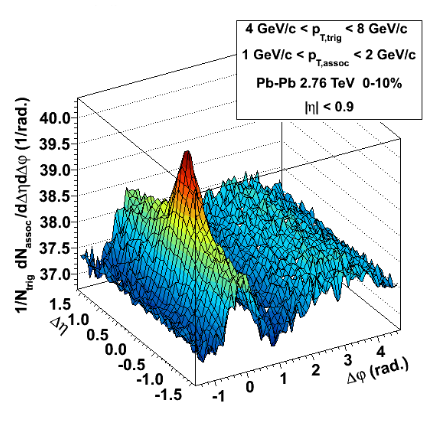}
\includegraphics[scale=0.14]{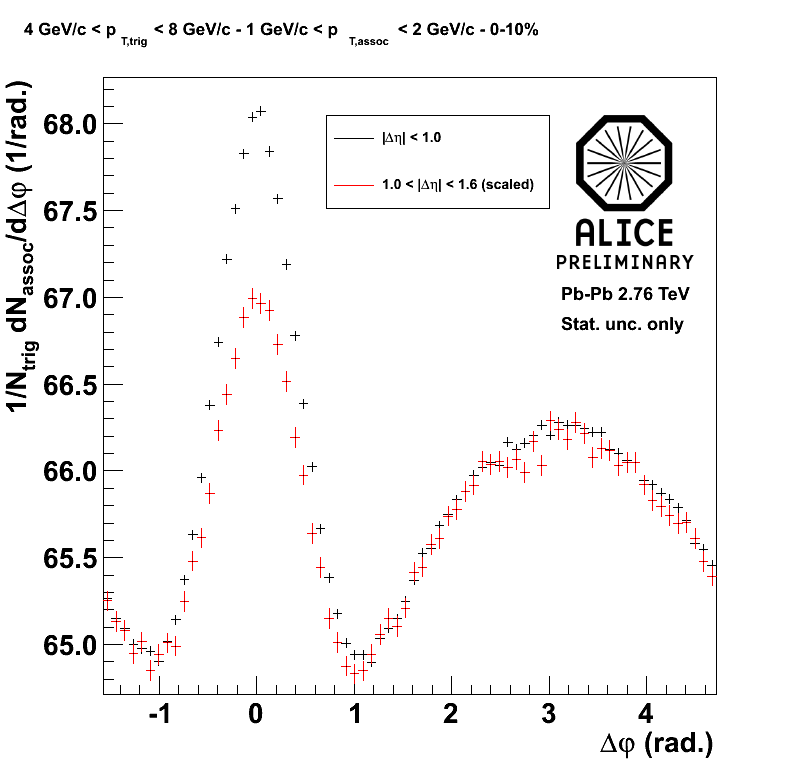}
\includegraphics[scale=0.28]{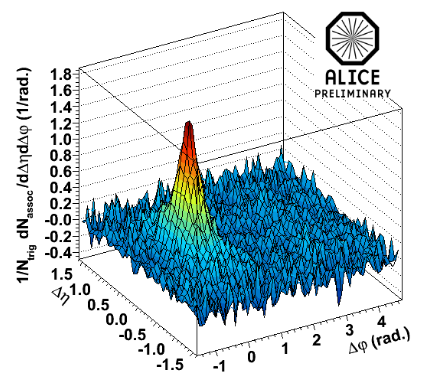}
\caption[]{\label{nearsideshape}Left: per-trigger yield; middle: per-trigger yield projection to $\Delta\phi$ in 
$1 < |\Delta\eta| < 1.6$ (red) and $|\Delta\eta| < 1$ (black); right: per-trigger yield subtracted flow contributions.
Shown is at trigger $4 < \ptt < 8~\gmom$, associated $1 < \pta < 2~\gmom$ in most central Pb-Pb collisions\cite{Jan1}.}
\end{figure}

A typical per-trigger yield is shown in the left panel of Fig.~\ref{nearsideshape}. At low $\pt$, per-trigger yield includes a 
sizable contribution from collective flow with a strong modulation in $\Delta\phi$ but independent of $\Delta\eta$. For isolating 
jet-like correlations to study the shape of the near-side jet peak, the flow contributions are determined in the long-range 
correlation region at $1 < |\Delta\eta| < 1.6$ and subtracted from the short-range correlation region at $|\Delta\eta| < 1$.
This prescription called the $\eta$-gap method provides a measurement independent of the flow strength.
The middle panel of Fig.~\ref{nearsideshape} shows the projection to azimuthal $\Delta\phi$ in $1 < |\Delta\eta| < 1.6$ (red)
and $|\Delta\eta| < 1$ (black). The difference between the two distributions in the near-side is the signal to be searched.
The away-side peak is removed by construction in this procedure. Hence, the away-side region can not be studied with this method.
The right panel of Fig.~\ref{nearsideshape} shows the subtracted per-trigger yield distribution in $\Delta\phi$ and $\Delta\eta$ 
with $4 < \ptt < 8~\gmom$ and $1 < \pta < 2~\gmom$ in most central Pb-Pb collisions.

\begin{figure}
\centering
\includegraphics[scale=0.22]{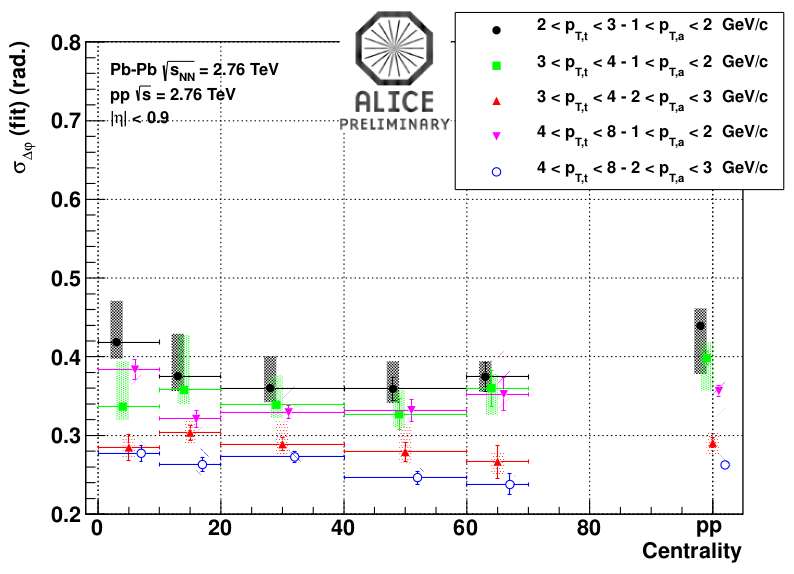}
\includegraphics[scale=0.22]{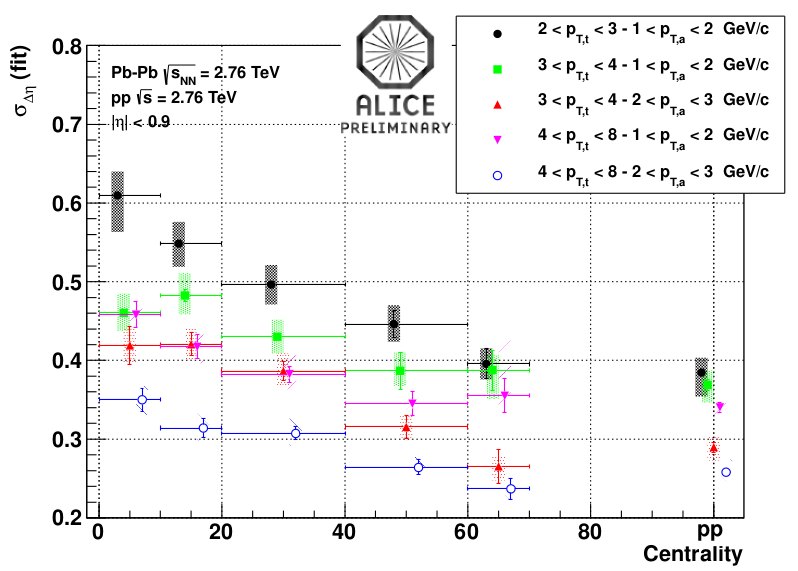}
\caption[]{\label{sigmaphieta}Centrality dependence of $\sigma_{\Delta\phi}$ (left) and $\sigma_{\Delta\eta}$ (right) in
five different $\ptt$ and $\pta$ $\pt$ bins.}
\end{figure}

In order to quantify the near-side peak shape, the peak is fitted with a sum of two 2D Gaussians with the center at
$\Delta\phi = \Delta\eta = 0$. The fit parameters are used to calculate the $rms$ (equal to the square root of the variance,
$\sigma$, for distributions centered at 0) in $\Delta\phi$ and $\Delta\eta$ direction ($\sigma_{\Delta\phi}$, $\sigma_{\Delta\eta}$). 
Fig.~\ref{sigmaphieta} presents the centrality dependence of $\sigma_{\Delta\phi}$ and $\sigma_{\Delta\eta}$ together with
reference results from pp collisions in five different bins of $\ptt$ and $\pta$. The results indicate that 
the $\sigma_{\Delta\phi}$ is independent of centrality within the errors, and decreases with increasing $\ptt$ 
and $\pta$, whereas the $\sigma_{\Delta\eta}$ has a significant increase of moving from pp to central collisions 
and also decreases with higher $\ptt$ and $\pta$. More details about this analysis can be found 
in\cite{Jan1,Jan2}.

\subsection{Modification of the jet-particle yield}
\label{modificationfactor}
At higher $\pt$ ($\ptt > 8~\gmom$, $\pta > 3~\gmom$) where collective effects are small and jet-like 
correlations dominate, the medium modification of the jet-particle yield has been studied by calculating ratios of yields
on the near-side and away-side. In order to remove uncorrelated background from the yield, a pedestal value is determined
by a constant fitting the region close to the minimum of the $\Delta\phi$ distribution ($\Delta\phi \approx \pm \frac{\pi}{2}$)
where uncorrelated background is dominated. A background shape considering the elliptic flow parameter $v_{2}$ is also analyzed.
For a given $\pt$ bin, the $v_{2}$ background is calculated 
as $2\langle v_{\rm 2,trig} \rangle \langle v_{\rm 2,assoc} \rangle \cos 2\Delta\phi$. The $v_{2}$ values are taken from an 
independent measurement\cite{ALICEflow}. The $\eta$-gap method, described in Sec.~\ref{nearsidejetshape},
is also used to remove the contributions from $\Delta\eta$-independent correlations on the near-side of the per-trigger yield. 
Subsequent to the background subtraction, the near-side and away-side yields are integrated within $|\Delta\phi| < 0.7$ and
$|\Delta\phi \pm \pi| < 0.7$, respectively.

\begin{figure}
\centering
\includegraphics[scale=0.6]{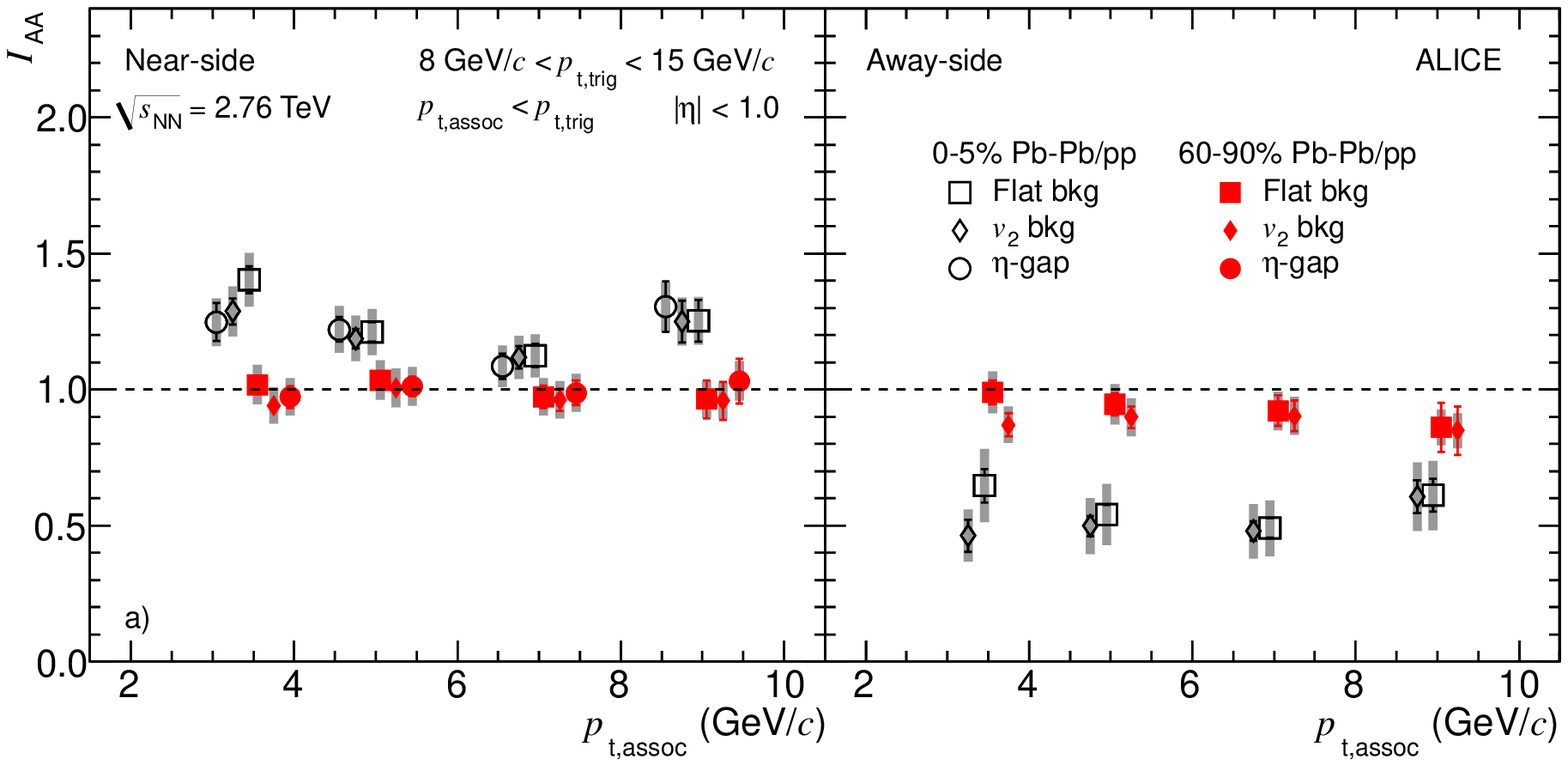}
\includegraphics[scale=0.6]{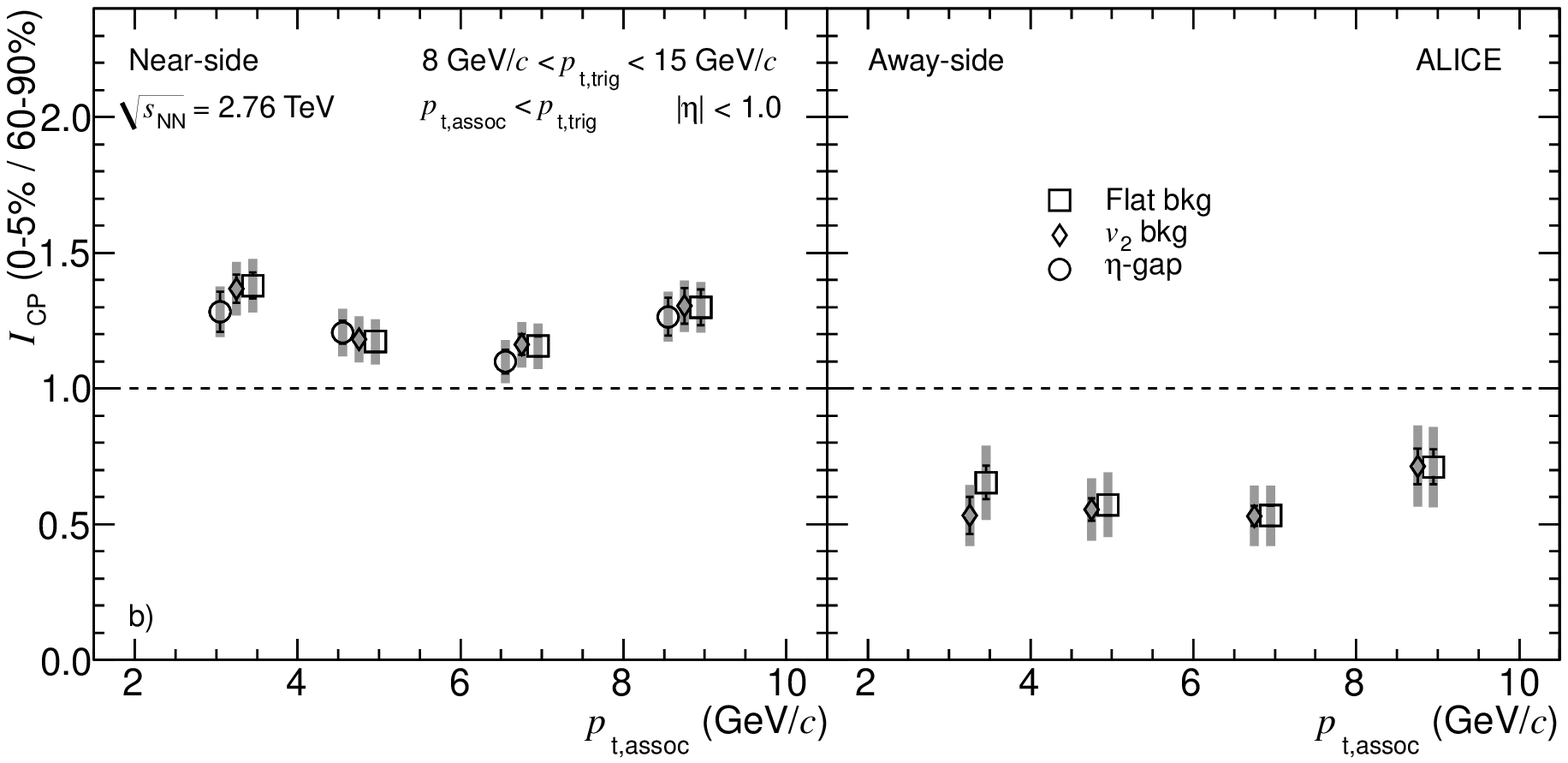}
\caption[]{\label{IAAICP} $I_{\rm AA}$ (top panel) for central (open black symbols) and peripheral (filled red symbols) collisions
, and $I_{\rm CP}$ (bottom panel). Different background subtraction schemes, a flat pedestal (squares), $v_{2}$ subtraction 
(diamonds) and $\eta$-gap subtraction (circles, only near-side) are presented\cite{IAAPaper,Jan3}.
}
\end{figure}

The modification of the jet-particle yield is calculated by the ratio of the per-trigger yield in Pb-Pb to pp collisions 
($I_{\rm AA}$) and the yield in central to peripheral in Pb-Pb collisions ($I_{\rm CP}$) with $I_{\rm AA} = Y_{\rm Pb-Pb}/Y_{\rm pp}$ 
and $I_{\rm CP} = Y_{\rm central}^{\rm Pb-Pb}/Y_{\rm peripheral}^{\rm Pb-Pb}$, respectively. The top panel in Fig.~\ref{IAAICP} presents
the yield modification factor $I_{\rm AA}$ for central and peripheral Pb-Pb collisions using the three 
background subtraction schemes as discussed. The main significant difference is in the lowest $\pta$ interval that 
confirms the small bias due to flow anisotropies in this $\pt$ region. In central collisions, an away-side suppression 
from in-medium energy loss is observed ($I_{\rm AA} \approx 0.6$). Moreover, there is an enhancement above unity of 
($I_{\rm AA} \approx 1.2$) on the near-side which has not been observed with any significance at lower collision energies\cite{lowIAA}. 
In peripheral collisions, both near-side and away-side are consistent with unity. 

Furthermore, the bottom panel in Fig.~\ref{IAAICP} shows the ratio of the yield in central and peripheral collisions, $I_{\rm CP}$.
The result of $I_{\rm CP}$ is consistent with $I_{\rm AA}$ in central collisions with respect to the near-side enhancement and the away-side 
suppression. 

A significant near-side enhancement of $I_{\rm AA}$ and $I_{\rm CP}$ in the $\pt$ region observed shows that the near-side parton
is also subject to medium effects. $I_{\rm AA}$ is sensitive to (i) a change of the fragmentation function, (ii) a possible change
of the quark/gluon jet ratio in the final state due to the different coupling to the medium, and (iii) a bias on the parton
$\pt$ spectrum after energy loss due to the trigger particle selection. More details about this analysis can be found
in\cite{IAAPaper,Jan3}.

\subsection{Fragmentation function}
\label{fragmentationfun}
\begin{figure}
\centering
\includegraphics[width=6cm, height=5.85cm]{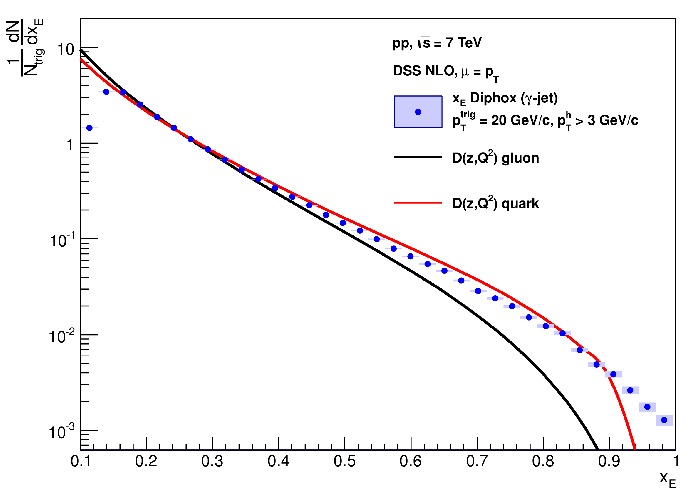}
\includegraphics[scale=0.30]{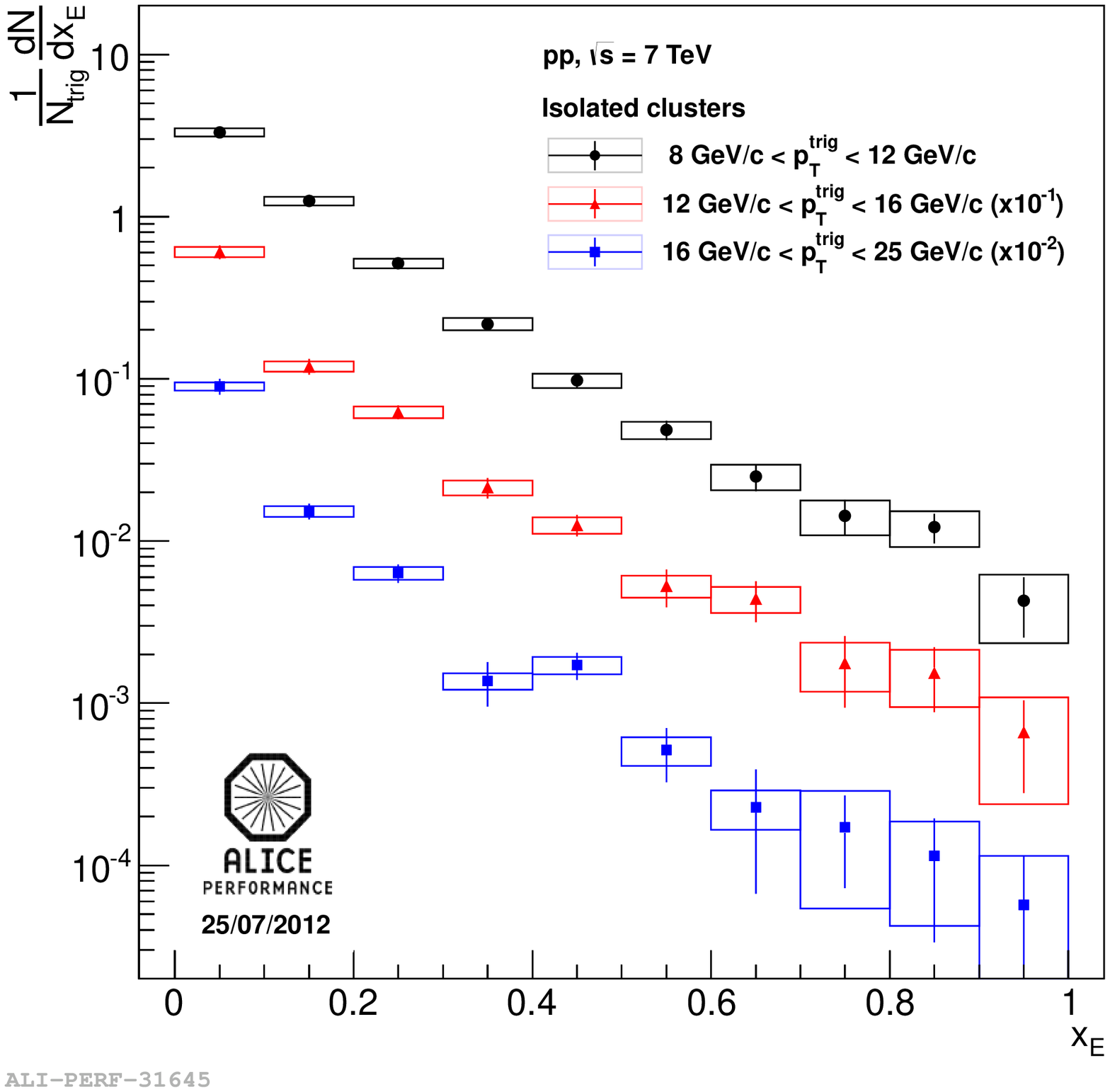}
\caption[]{\label{diphox} Left: $x_{\rm E}$ distribution from $\gamma$-jet production produced by Diphox, and compared to
DSS quark and gluon fragmentation; right: $x_{\rm E}$ distribution of isolated cluster (photon candidate)-hadron correlations
in three $\ptt$ bins\cite{ALICEIsoPhoton}.}
\end{figure}

In isolated photon-hadron correlations, the away-side distribution provides a measurement of the full fragmentation function of
the jet at the opposite azimuthal direction of the isolated photon. In leading order pQCD, the fragmentation function of the recoil
jet from the away-side parton should be given to a good approximation by the imbalance parameter $x_{\rm E}$ distribution as:

\begin{equation}
x_{\rm E} = - \frac{\vec{p}_{T}^{\gamma}\cdot\vec{p}_{T}^{h^{\pm}}}{|\vec{p}_{T}^{\gamma}|^{2}} = 
-\frac{|\pt^{h^{\pm}}|\cos\Delta\phi}{|\pt^{\gamma}|}
\end{equation}
where $\Delta\phi$ is the azimuthal angle between isolated photons and hadrons. 
The transverse and longitudinal momenta of away-side parton does not exactly balance with the isolated photon. Hence,   
the parameter $x_{\rm E}$ is an approximation rather than an exact measurement to the fragmentation function of the away-side
jet\cite{PHENIXFF}. This analysis is only performed with pp collisions at $\sqrt{s} = 7~\tev$. 
The left panel in Fig.~\ref{diphox} shows the $x_{\rm E}$ distribution computed from Diphox $\gamma$-jet 
production\cite{Diphox} and comparison with DSS quark and gluon fragmentation function\cite{DSS}. It indicates that the 
$x_{\rm E}$ distribution mainly follows the quark fragmentation behaviour in a large range (0.2 to 0.8) because of the 
dominant contribution of compton scattering process. 

\begin{figure}
\centering
\includegraphics[scale=0.3]{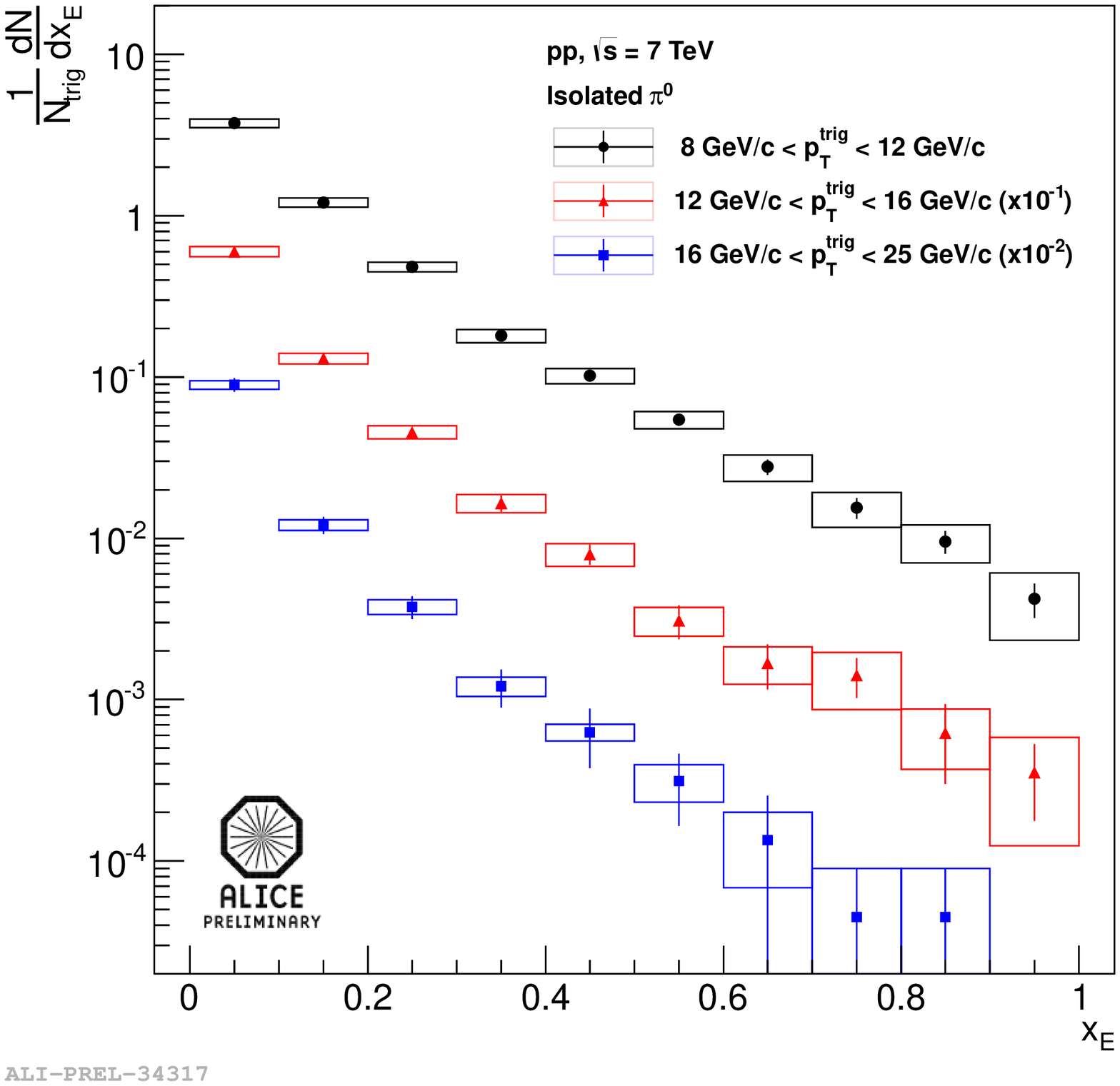}
\includegraphics[scale=0.3]{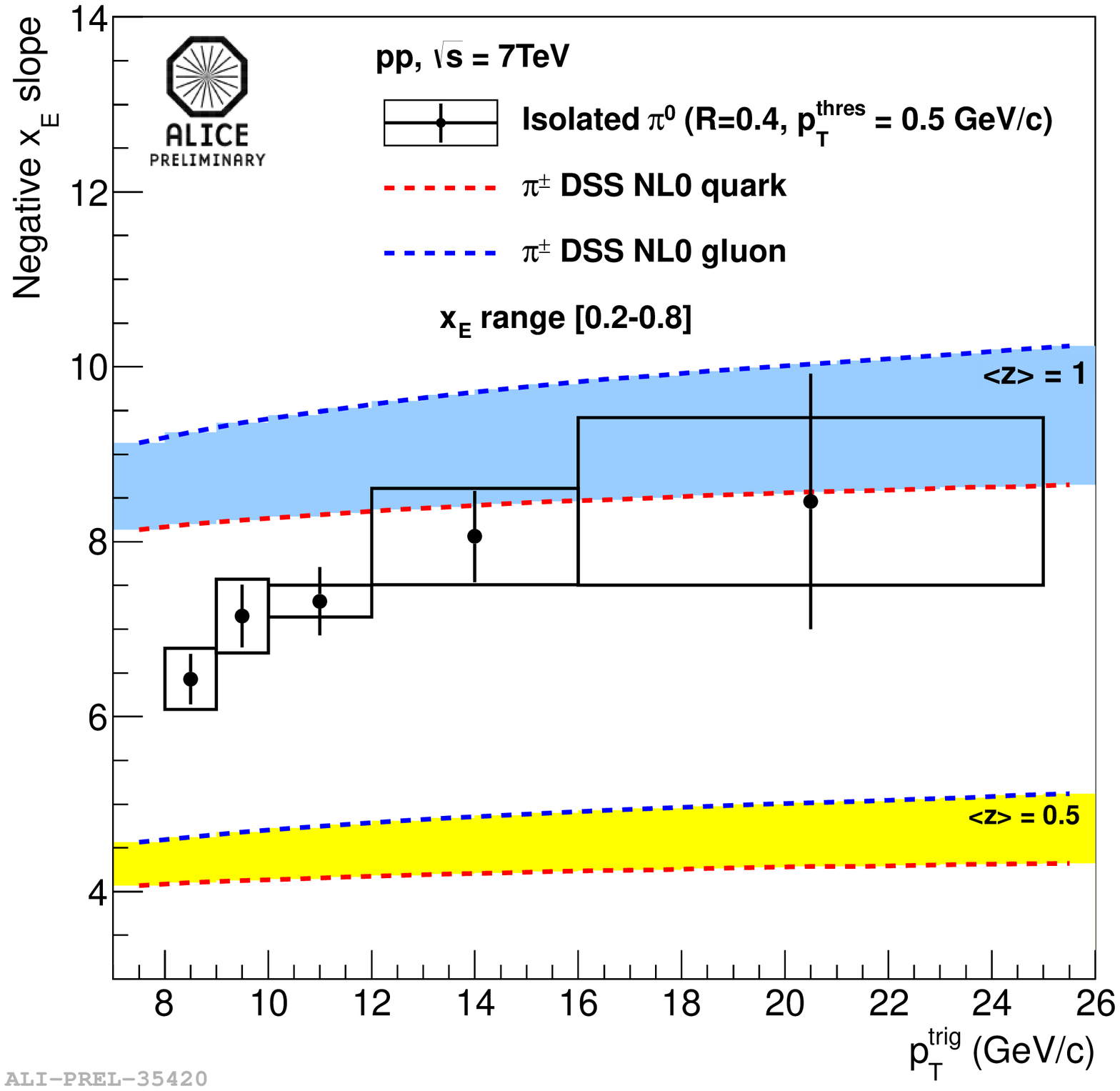}
\caption[]{\label{pi0xE}Left: $x_{\rm E}$ distributions of isolated $\pi^{0}$-hadron correlations in three $\ptt$ bins; 
right: slopes extracted from exponential fit of isolated $\pi^{0}$-hadron correlations and compared to DSS quark-gluons
fragmentation functions\cite{ALICEIsoPhoton}.}
\end{figure}

Clusters filtered by EMCal photon identification cuts, photon candidates, are dominated by a large fraction of decay photons of neutral
mesons (mostly $\pi^{0}$). The fraction is reduced about 80\% by applied isolation criteria. In this analysis, the isolation
criteria requires no particles including charged and neutral particles with $\pt > 0.5~\gmom$ in a cone of radius 
$R=\sqrt{\Delta\phi^{2}+\Delta\eta^{2}} = 0.4$ around a photon candidate with largest $\pt$ in one event. The right panel in
Fig.~\ref{diphox} presents the $x_{\rm E}$ distribution from isolated cluster (photon candidate)-hadron correlations with isolated 
leading cluster $\pt$ at $8 < \ptt < 12~\gmom$ (black), $12 < \ptt < 16~\gmom$ (red), and $16 < \ptt < 25~\gmom$ (blue).
Two decay photons from high $\pt$ $\pi^{0}$ are generally close and their two electromagnetic showers overlapping in the calorimeter cells
are clustered. A fraction of the clusters rejected unsuccessfully by photon identification cuts are the dominant contamination 
of isolated photons. In order to subtract the contamination, the $x_{\rm E}$ distribution of isolated $\pi^{0}$-hadron correlations
is measured, see the left panel in Fig.~\ref{pi0xE}. Compared to inclusive $\pi^{0}$, the isolated $\pi^{0}$ equally 
carries a large fraction of its parent parton energy from 0.5 to 0.8. An exponential slope is extracted from fitting 
the $x_{\rm E}$ distribution of isolated $\pi^{0}$-hadron correlations and compared to DSS fragmentation functions shown in 
the right panel in Fig.~\ref{pi0xE}. The comparison indicates that the isolated $\pi^{0}$ is a parton fragmentation product 
and $\pt^{\pi^{0}} < \pt^{\rm parton}$.

\begin{figure}
\centering
\includegraphics[scale=0.3]{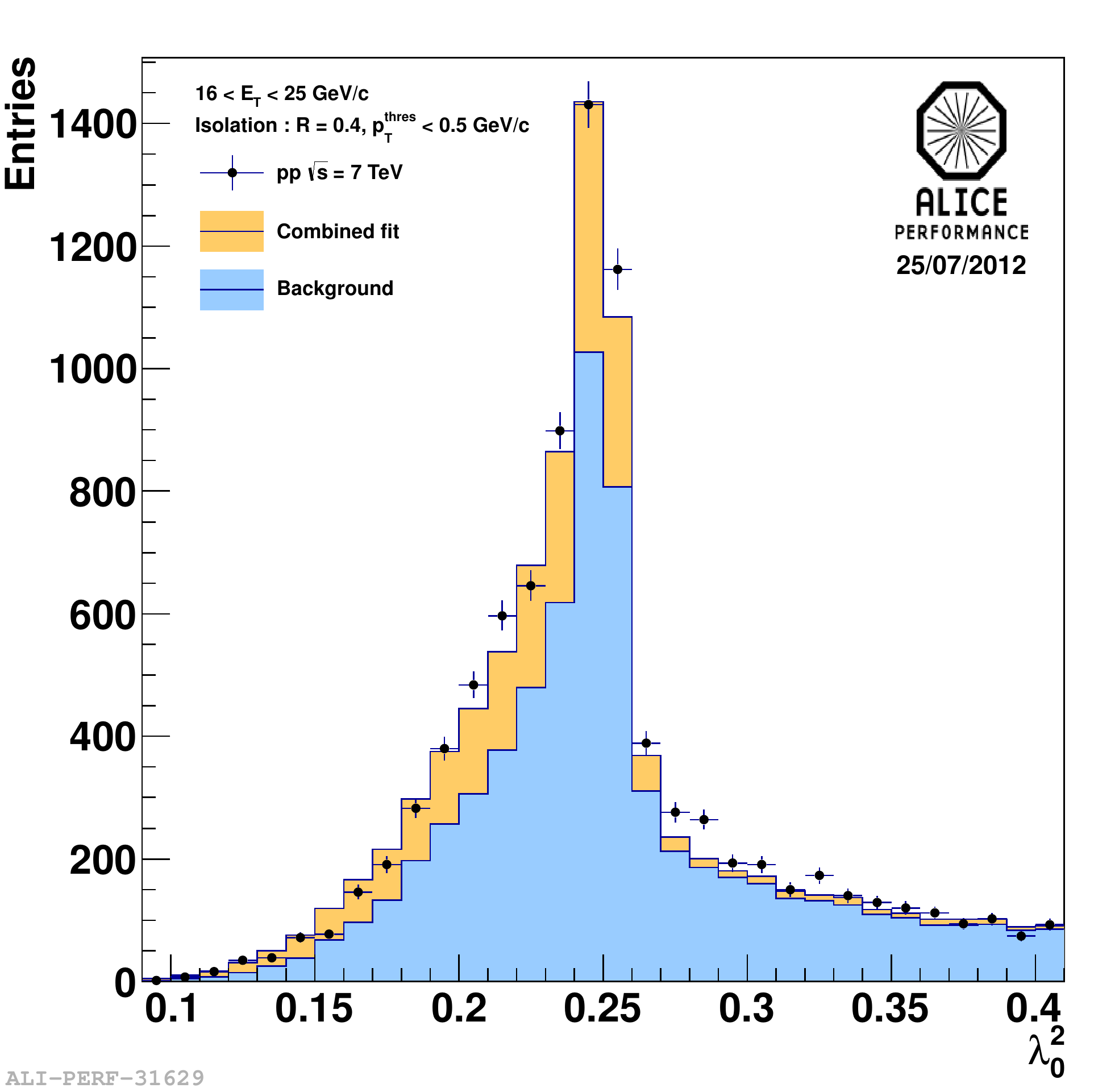}
\includegraphics[scale=0.3]{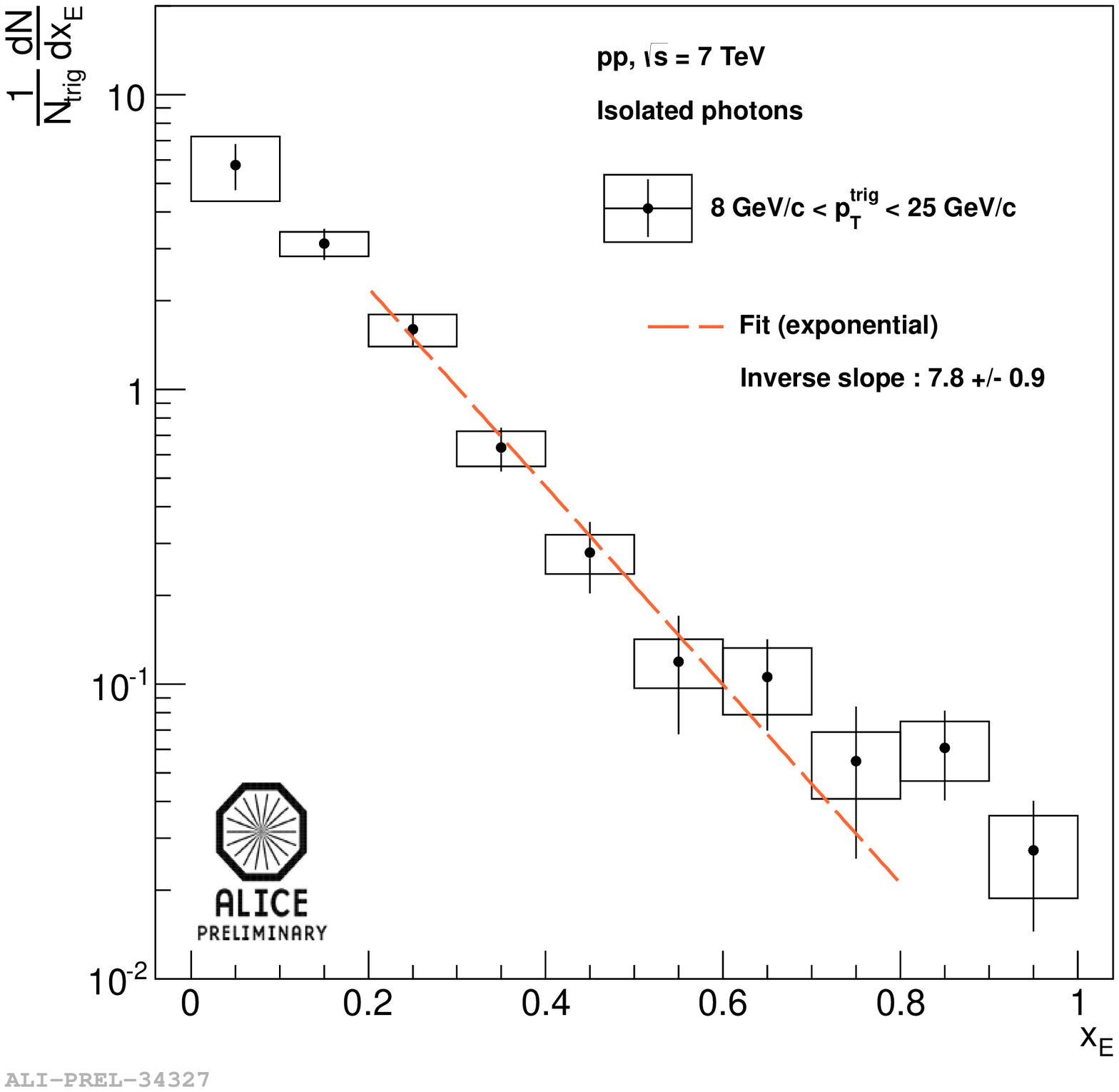}
\caption[]{\label{islaotedPhoton}Left: isolated cluster shower shape long axis $\lambda_{0}^{2}$ distribution fitted by a
two-component binned likelihood; right: $x_{\rm E}$ distributions of isolated photon-hadron correlations at 
$8 < \pt^{\rm iso~\gamma} < 25~\gmom$\cite{ALICEIsoPhoton}.}
\end{figure}

To subtract the contamination contribution to the $x_{\rm E}$ distribution, the isolated photon purity is estimated firstly by
two-component binned likelihood method: a mix of scaled signal and contamination distribution is used to fit all clusters in pp 
collision data at shower shape long axis $\lambda_{0}^{2}$ distribution, see the left panel in Fig.~\ref{islaotedPhoton}. Here,
the signal component is obtained from $\gamma$-jet events generated with PYTHIA and propagated through the detectors with
GEANT3, and the contamination component is extracted from data by selecting events which have failed the isolation 
criteria. The typical purity values obtained from this method in $8 < \pt < 25~\gmom$ increase from about 5\% to 70\%. 
The $x_{\rm E}$ distribution of isolated $\pi^{0}$-hadron correlations scaled with respect to the isolated photon purity
estimated previous is subtracted from isolated cluster-hadron correlations. In the meanwhile, the underlying events $x_{\rm E}$
contributions which are estimated at two different regions $\frac{\pi}{3} < \Delta\phi < \frac{2\pi}{3}$ and 
$\frac{4\pi}{3} < \Delta\phi < \frac{5\pi}{3}$ are also removed from the isolated cluster-hadron correlations. The $x_{\rm E}$ 
distribution of isolated photon-hadron correlations is shown in the right panel in Fig.~\ref{islaotedPhoton}, and a slope 
$7.8 \pm 0.9$ is obtained from the fitting of $x_{\rm E}$ distribution at $0.2 < x_{\rm E} < 0.8$.  More details about this analysis
can be found in\cite{ALICEIsoPhoton}. 

\section{Summary}
\label{summary}
Two-particle correlations have been used to study the properties of the hot and dense medium with ALICE at LHC. In di-hadron correlations,
the medium effect on the near-side jet peak is quantified at transverse momenta below $10~\gmom$. A broadening at lower $\pt$
intervals of trigger and associated particles and in more central Pb-Pb collisions is observed. The near-side peaks show a 
significant increase in $\Delta\eta$ moving from pp to central Pb-Pb collisions and no centrality dependence in $\Delta\phi$ within
errors. This might be an indication of interaction of jets with longitudinal flow. At higher $\pt$, the modification factors 
$I_{\rm AA}$ and $I_{\rm CP}$ of the jet-particle yield show a strong suppression on the away-side consistent with strong 
medium energy loss as well as an interesting near-side enhancement by effect of medium at the LHC. The fragmentation function
in pp collisions is calculated by the imbalance parameter $x_{\rm E}$ extracted from isolated leading photon-hadron 
correlations. 

\section{Acknowledgement}
This work is partly supported by the ``973'' Grant of MOST of China 2013CB837803, the NSFC Key Grant 11020101060, 
IRG11221504, 11375071, 11005044, the CCNU Key Grant CCNU13F026 and QLPL2012P01.


\end{document}